# Modeling Subsurface Charge Accumulation Images of a Quantum Hall Liquid

S.H. Tessmer[a], G. Finkelstein[b,†], P.I. Glicofridis[b], and R.C. Ashoori[b],

[a]*Department of Physics and Astronomy, Michigan State University,
East Lansing, MI 48824*

[b]*Department of Physics, Massachusetts Institute of Technology,
Cambridge, Massachusetts 02139*

[†]*Current Address: Physics Department, Duke University, Box 90305,
Durham, NC 27708*

Subsurface Charge Accumulation imaging is a cryogenic scanning probe technique that has recently been used to spatially probe incompressible strips formed in a two-dimensional electron system (2DES) at high magnetic fields. In this paper, we present detailed numerical modeling of these data. At a basic level, the method produces results that agree well with the predictions of models based on simple circuit elements. Moreover, the modeling method is sufficiently advanced to simulate the spatially resolved measurements. By comparing directly the simulations to the experimentally measured data, we can extract quantitatively local electronic features of the 2DES. In particular, we deduce the electron density of states inside the incompressible strips and electrical resistance across them.

PACS numbers: 73.40.Hm, 73.23.-b, 7320.Dx, 73.23.Ps

## I. INTRODUCTION

Recently developed scanning probe microscopy methods based on electric-field sensing provide a new window into the quantum mechanics of confined systems and may play a significant role in characterizing future nano-electronic devices.[1-4] In particular, Subsurface Charge Accumulation (SCA) imaging is a cryogenic technique that locally measures the accumulation of mobile charges within a conducting system.[5] By rastering the probe over the surface, we can generate images that show the spatial patterns of the charging – even if the conducting layer is buried beneath a 100 nm thick dielectric. Rather than driving a transport current through the conductor, which requires the application of a potential across it, we apply a single AC potential – as indicated in Fig. 1. Due to capacitive coupling, the excitation causes charge to flow in and out of the interior at the applied frequency $f$. This results in AC image charge at the apex of the probe, which we detect the using a circuit constructed from field-effect transistors mounted in close proximity to the tip.[6]

This paper focuses on modeling recent SCA measurements of a two-dimensional electron system (2DES). [7] The application of a perpendicular magnetic field can cause parts of the interior of the 2DES to act as an

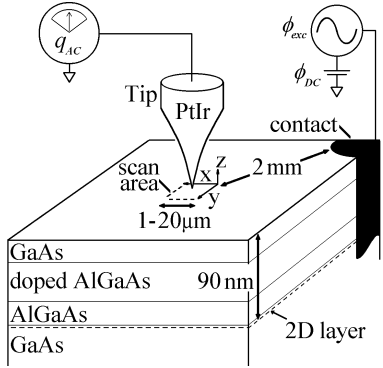

Fig. 1. Schematic of the SCA measurement and sample geometry. The applied AC excitation $\phi_{exc}$ causes charge to flow in and out of the 2D layer. This couples capacitively to the tip resulting in AC image charge, which we detect the using a circuit constructed from field-effect transistors.

insulator – a phenomenon intimately connected to the quantum Hall effect. Many of the key features observed in the quantum Hall effect may be explained in terms of the transport through the quasi-one-dimensional edge channels.[8] Each channel is formed where the energy of the corresponding Landau level at the edge of the sample equals the Fermi energy. Theories predict that the edge channels should be separated by narrow strips with precisely integer Landau level filling.[9,10] In the model, the strips arise due to the low compressibility of the 2DEG at the cyclotron gap in the electron density of states between Landau levels. Our SCA measurements immediately yield the spatial extent of the charging patterns. However, to quantitatively extract local conductivity and compressibility information, these data must be carefully compared to calculations that account for the sample charging and the measurement technique itself.

The sample used for these measurements was an $Al_{.3}Ga_{.7}As$ / GaAs wafer grown by molecular beam epitaxy. The 2DES forms from electrons that become trapped at the GaAs / AlGaAs interface 90 nm below the surface. The average electron density in this layer was $n_b = 3.5 \times 10^{11}$ cm$^{-2}$ and the transport mobility was approximately $4 \times 10^5$ cm$^2$/Vs. The measurements were performed with the microscope immersed in liquid helium-3 at a temperature of 300 mK. Most of the electric field lines that terminate on the tip electrode do so on the portions far from the sharp apex. Hence we have a large background signal that adds an offset to the signal of interest, which we subtract away using a bridge circuit.[6] To minimize the DC perturbation of the tip, we typically fix the potential at $\phi_{DC} = 0.8$ V. This compensates for the tip-sample contact potential [11], eliminating DC electric fields between the tip and 2DES.

In general, both the conducting layer's self-capacitance and the conductor-to-tip mutual-capacitance will contribute to the magnitude of the charging signal. A local suppression of the signal may result from either low compressibility or low conductivity. The compressibility (i.e. density of states) is given by $D = dn/d\mu$. $D$ sets the amount of charge that will accumulate in the 2DES for a given potential variation in the DC limit. With regard to the conductivity, RC charging time effects may result in insufficient time to charge the region compared to the period of the AC excitation. As the 2DES longitudinal conductivity is very low in the vicinity of integer filling, this possibility must be examined carefully. To distinguish between the two mechanisms we study both the in-phase signal $Q_{in}$ and the 90° out-of-phase (lagging) component $Q_{out}$ as a function of frequency. Therefore, we obtain two images simultaneously with each scan.

## II. SIMPLE MODELS

To better understand our modeling scheme, presented in the next section, it is useful to first introduce simple models of the charging. Moreover, these models will provide basic tests for the advanced method.

## A. Single Parallel Plate Capacitor

The induced charge on the tip is closely related to the local charging of the 2DES, as the tip will capture more field lines when it is above a region of relatively high charge accumulation. The simplest model of our system is a single parallel plate capacitor, as shown in Fig. 2(a). Although this is a severe simplification, if the 2DES is sufficiently conducting so that charging delays can be neglected (at a given frequency), the single capacitor model can be used to crudely estimate the compressibility contribution to the signal. Here the applied excitation potential is assumed as a constant across the sample, and the charge induced on the tip is set exclusively by the tip-sample mutual capacitance $C(D)$, to which both the geometry and the compressibility contribute. Using MKS units $C(D)$ takes the following form:

$$C(D) \propto \left( \frac{h}{\varepsilon_0} + \frac{d}{\kappa \varepsilon_0} + \frac{1}{e^2 D} \right)^{-1}, \quad (1)$$

where $h$ is the effective distance from the surface to the tip, $d$ is the 2DES depth below the surface, and $\kappa$ is the dielectric constant of the semiconductor.[12] Finally, the charge on the tip is simply the product of the capacitance and the excitation potential, $q_{AC} = \phi_{exc} C(D)$.

## B. RC Circuit

To describe systems for which either the bulk conductivity or local conductivity is sufficiently low to result in significant charging delays, the above model must be extended – as the magnitude and phase of the potential will vary with position. Hence, for more advanced models we introduce the local

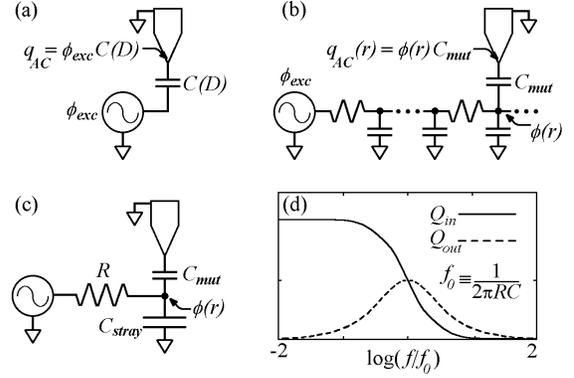

Fig. 2. Simple models for the system and corresponding expressions for the measured charge. (a) If RC charging delays are negligible, a single parallel-plate-capacitor model provides an estimate for the compressibility contribution ($D$) to the signal. (b) If the impedance of the 2D layer gives significant delays, the measurement probes the local potential $\phi(r)$, which will have a phase shift and reduced magnitude relative to $\phi_{exc}$; $C_{mut}$ is the tip-2D mutual capacitance. In these cases, a simple RC model, shown in (c), provides an intuitive picture of the charging. The sample is modeled as a single resistor $R$ charging two capacitors to ground: $C_{mut}$ and $C_{stray}$, where $C_{stray}$ accounts for electric flux emanating from the 2DES that does not terminate on the tip. (d) Charging characteristics of the RC model as a function of the excitation frequency $f$. Here $C = C_{mut} + C_{stray}$, and $Q_{in}$ and $Q_{out}$ are in the in-phase and out-of-phase components of the charging, as described in the text.

effective potential $\phi(r)$. With the tip centered above $r$, we can then write the charge induced as

$$q_{AC}(r) = \phi(r) C_{mut}, \quad (2)$$

where $C_{mut}$ is the mutual capacitance between the tip and the 2D layer.

Fig. 2(b) shows the basic picture for charging the tip. As the potential propagates across the sample, its magnitude is reduced and phase shifted by the resistance and self-capacitance of the system along the path. Finally, the point below the tip with potential $\phi(r)$ determines the measured charge, according to Eq (2). We can explicitly define the measured charging components $Q_{in}$ and $Q_{out}$ using complex notation for the potentials: $\phi_{exc}=\phi_{exc0}\, e^{\,i\omega t}$ and $\phi(r)=\phi_0(r)e^{\,i(\omega t+\delta)}$, where the injection frequency is $f=\omega/2\pi$ and $\delta$ is the local phase delay. Defining the complex charging per unit excitation voltage as $Q = (\phi(r)/\phi_{exc})\, C_{mut}$, the two components are given by $Q_{in} = Re\,\{Q\}$, and $Q_{out} = -Im\,\{Q\}$.

Figures 2(c) and 2(d) present a simplified view of the RC system, which in many respects provides the correct picture for the conductivity contribution to the charging. We consider the sample as consisting of a single resistor $R$ charging two capacitors to ground: $C_{mut}$ and $C_{stray}$, where $C_{stray}$ accounts for electric flux emanating from the 2DES that does not terminate on the tip. Fig. 2(d) shows the $Q_{in}$ and $Q_{out}$ components of tip charge for this model, where $C = C_{mut} + C_{stray}$. With respect to the in-phase charging, at frequencies greater than the roll-off frequency $f_0 = 1/2\pi RC$, the excitation cycle is too short for full charging to occur. As a result, $Q_{in}$ diminishes monotonically with increasing $f$. In contrast, the out-of-phase signal $Q_{out}$ displays a peak at $f_0$. This behavior is easy to understand. At low frequencies, charging occurs rapidly compared to the period of the excitation, so the charge signal is in-phase with the excitation, and $Q_{out}$ is small. At intermediate frequencies, the charge lags the excitation, and $Q_{out}$ is large. At high frequencies, little charge enters the sample, so $Q_{out}$ and $Q_{in}$ both tend to zero.

## III. MEASUREMENTS AND ADVANCED MODELING

Fig. 3 (a) shows a SCA image of a ring-shaped feature of reduced charging that appeared at applied magnetic fields near 4 T. At the center of this area the sample was prepared to have a density maximum, resulting in an electron density gradient of $dn/dx \approx 5 \times 10^{10}/cm^2\mu m$ near the edges of the ring.[7] The dark feature marks the density contour near integer filling; in this case four spin-split Landau levels are filled ($\nu=4$). As the field increases by a small increment, integer filling occurs at slightly higher densities. Indeed, the ring feature was observed to move up the density gradient and hence to shrink in diameter as expected. Fig. 3(c) (left column) presents a series of measurements as the tip scanned across a single line through the center of the ring. Both $Q_{in}$ and $Q_{out}$ curves are plotted at four magnetic fields and at three different frequencies, 10 kHz, 30 kHz, and 100 kHz. We see that $Q_{in}(x)$ shows the greatest changes with respect to $x$ at 4.3 T, where it is most reduced in the interior. In contrast, $Q_{out}(x)$ shows the most structure at 4.1 T and 4.2 T. The data were normalized with respect to the in-phase curve at 4.3 T; this choice is sensible in the view that the interior signal is likely near zero, i.e. no charge entering, whereas the exterior most likely represents full charging. The advanced modeling presented below supports this assumption.

In addition to $\nu=4$, the ring also appeared at filling factors $\nu=2$ and $\nu=6$ (not shown). The observed widths of the rings for $\nu=2$ and 4 were ~0.6 μm and ~0.4 μm respectively. These widths are about three times greater than predicted by the standard theoretical picture.[10] This discrepancy was the main focus in Ref. [7]. Disorder due to density fluctuations of charged donors in the doped layer of the sample represents the most likely explanation.[13,14] Effectively, these fluctuations give rise to states within the gap between Landau levels, with the resulting increase in screening yielding wider strips.

Two physical phenomena can cause structure in the local charging, both of which are expected to occur near integer filling: (1) a decrease in the local compressibility, $D= dn/d\mu$, and (2) a decrease in the local conductivity $\sigma$. Much of the data of Fig. 3 shows out-of-phase structure, indicating that the 2DES does not have enough time to fully charge during the excitation cycle. Hence in addition to compressibility variations, changes in conductivity must contribute significantly. To untangle the effects of these two phenomena, we have developed a numerical method that models the charging in two dimensions within the 2D layer – generating calculations that can be compared directly to the measurements.

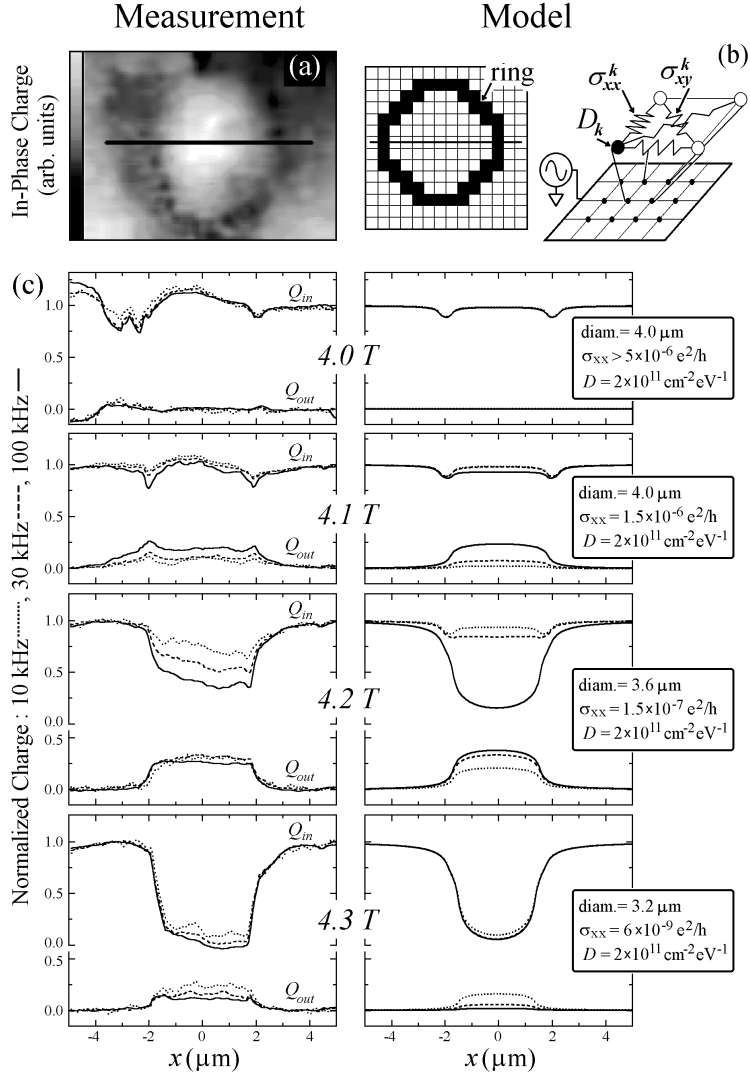

Fig 3. (a) SCA image in the vicinity of a density maximum at a magnetic field of 4.05 T. The density contour corresponding to $\nu=4$ forms a dark ring for which the charging is reduced. To investigate the charging characteristics in detail, numerous measurements were acquired by scanning the tip along the indicated line. (b) Schematic of the advanced method used to model the data. (c) Measurements (left) and best-fit modeling curves (right) of the ring feature as a function of frequency and magnetic field. The insets show the parameters of the respective calculations, as described in the text.

An important key to understanding the behavior of the system is the ratio of $C_{mut}$ to $C_{stray}$. Consider the charging of a region of the 2DES of radius *L*, situated directly below the probe. Electric field lines connect the charge within this region to the tip. If the tip is sharp and narrow compared to *L*, the field lines terminate on the bottom up to a height of ~*L* above the apex. The result is a mutual capacitance of approximately $C_{mut} \approx 2\pi\varepsilon_0 L$.[15] For the PtIr tips used in this study, the nominal radius of curvature was ~50 nm with a cone angle of ~10 degrees.[16] So for micron-sized sample features, we should indeed be in the sharp-tip limit. With respect to the stray capacitance, we can consider the self-capacitance of a disk-shaped region of radius L near the surface of a semi-infinite dielectric slab: $C_{self} \approx 8(\kappa+1)\varepsilon_0 L/2$. Comparing this to $C_{mut}$, we see the self-capacitance is roughly a factor of $\kappa$ greater than the mutual capacitance. Therefore due to the high permeability of GaAs, $\kappa=12.5$, we expect $C_{self} \gg C_{mut}$. Lastly, in the limit of negligible mutual capacitance, $C_{stray}=C_{self}$. Hence we can conclude that $C_{stray} \gg C_{mut}$ in the sharp-tip limit.

The roll-off frequency represents another key parameter. In contrast to the simple one-dimensional network of resistors and capacitors, the realistic model of the 2DES must account for the distributed *RC* system that comprises the layer. In general, a distributed *RC* system differs from the model shown in Fig. 2(c) and 2(d) in that there is no unique roll-off frequency. The roll-off frequency to charge a particular region depends on its size and average resistance. For example, we can again consider a disk-shaped region of radius *L* with effective resistance $R_L$. To simplify the example, we assume that the surrounding 2DES is of arbitrarily low impedance so that the injected potential is effectively applied right at the edge. As the self-capacitance is proportional to the lateral dimension, we see that the roll-off frequency to charge this region is $f_0 = 1/2\pi RC \propto 1/R_L L$. So smaller length scales charge more quickly. Even for a large resistance, if $L \ll 1/\kappa\varepsilon_0 f_0 R_L$, the region will charge effectively over the period of the excitation.

We model the 2DES as a two-dimensional array of grid points, each with a distinct compressibility and conductivities to neighboring points. Following the above discussion, we assume $C_{stray} \gg C_{mut}$. In this limit it is reasonable to first calculate the complete charging within the interior of the 2DES due exclusively to its self-capacitance and electronic structure. This approach eliminates the need to repeat the time consuming calculations for every position of the tip. As a last step, we find the resulting charge induced on the tip due to $C_{mut}$; here we include the spatial resolution of the probe as discussed below. By repeating the procedure for numerous arrays of trial conductivity ($\sigma$) and compressibility (*D*) patterns, we can generate sets of calculations to be compared directly to the experimental measurements.

As described below, for a given geometry of a distinct low-compressibility feature, such as a ring, we find $\sigma$ and *D* contribute in qualitatively different ways to the spatial distribution of the charging. Hence if fits to the measurements can be found that reasonably reproduce the detailed structures, we can assert with some confidence that the parameters of the model represent an appropriate description of the actual system. In this way, quantitative local compressibility and conductivity information can be extracted.

Fig. 3 (b) schematically shows the parameters of our numerical routine, which calculates the effective AC potential $\phi(r)$ for every point above an *N* x *N* grid, resulting from an AC excitation of frequency *f* injected uniformly around the perimeter. In general, for each of $N^2$ points *k*, the compressibility $D_k$, as well as the longitudinal and Hall conductivities, $\sigma_{xx}^k$ and $\sigma_{xy}^k$, to neighboring points represent adjustable parameters. Given the dielectric properties of the semiconductor and the distance of the 2DES below the surface, the routine calculates the internal capacitance of the array. With the array's conductivity and capacitance set, it is straightforward to calculate the effective potential of every point of the array $\phi_k$ (i.e., $\phi(r)$), in analogy to the one-dimensional problem of Fig. 2(b). The typical way to model capacitance measurements takes the mutual capacitance term to account for the compressibility contribution (e.g. Eq. 1). However in our scheme, $C_{mut}$ is taken as a constant function to be included as the final step. Hence, as detailed in the appendix, we adjust the effective potential $\phi(r)$ to account for the 2DES compressibility. The result is mathematically identical to the standard approach.

As the final step in the calculation, we find the charge induced on the tip due to the tip-2DES mutual capacitance. Although 2DES potential directly below the tip predominantly determines the charge on the apex, more distant 2DES regions contribute significantly to the charge induced on other parts of the tip. Hence, we must generalize the simple expression $q_{AC}(r) = \phi(r)C_{mut}$. To do this, we define a capacitance function that varies according to sample location, $c(r-r_0)$. This function represents the geometric mutual capacitance between the 2DES and the tip per unit area of the 2D layer, for the tip positioned above location $r_0$. Convolution with the effective potential then gives the charge on the tip,

$$q_{AC}(r_0) = \int \phi(r)c(r-r_0)d^2r. \quad (3)$$

The function $c(r-r_0)$ is crucial as it determines the spatial resolution of the probe. We expect it to be a bell-shaped function peaked at the location of the apex, and of width roughly equal to the tip-2DES separation $H$. For example, taking the tip as a thin rod of constant thickness and with a radius of curvature smaller than $H$, the form of $c(r-r_0)$ is approximately $H/\sqrt{H^2+(r-r_0)^2}$. [15] In general, the mutual capacitance function depends on the exact shape, broadening with both the tip radius and the angle of the conical portion of the tip. For the modeling presented here, we have used a numerically calculated form based on a realistic tip geometry: half cone angle=10°, apex radius=50 nm.[17]

Fig. 3(c) compares the same measured data (left) to calculated curves generated by our routine (right). For the calculated curves, the scanned area was modeled as a 15 x 15 array of grid points. The 2DES parameters were chosen to be homogeneous, except for a ring of width one grid spacing = 400 nm. As an example, Fig. 3(b) shows the pattern used for the 4.0 T and 4.1 T curves. In order to make a direct comparison, the convolution with $c(r-r_0)$ was performed over a straight line across the diameter. As this final convolution is performed separately, we are free to choose a finer point spacing for $r_0$. Hence the calculated curves are smooth on the scale of 400 nm. The measured images dictated the ring diameters, and the experimental frequencies $f$ were applied. Moreover, for all the points not on the ring (interior and exterior), $\sigma_{xx}$ and $D$ were set arbitrarily high. Therefore, only the conductivity and compressibility parameters on the ring itself represented free variables, adjusted to achieve the best fit to the measured data. The $\sigma_{xx}$ and $D$ best-fit parameters are shown as insets. We found that the calculations were independent of the Hall conductivity $\sigma_{xy}$ for the ring geometery. This should come as no surprise, considering the similarity to the Corbino-disk geometry.

Comparing the measurements to the model, it is clear that our calculation successfully reproduced the general shapes of the curves and evolution with magnetic field. The fits were achieved by taking the modeled compressibility of the ring as $D(ring) = 2 \times 10^{11}$ cm$^{-2}$eV$^{-1}$ for all values of the field. However, it was necessary to use a range of values for $\sigma_{xx}(ring)$ to reproduce the evolution with field of the interior charging signal. At 4.0 T, the measurements show a high interior in-phase signal and almost no out-of-phase structure. This is consistent with an arbitrarily high $\sigma_{xx}(ring)$. By finding the minimum conductivity consistent with a negligible out-of-phase signal, we set a lower limit of $\sigma_{xx}(ring) > 5 \times 10^{-6} \ e^2/h$. As the field increases, the measurements show a reduced in-phase signal, with significant out-of-phase structure. Our model implies this behavior stems from a drastically reduced conductivity across the ring. At 4.3 T, the best fit was achieved with $\sigma_{xx}(ring) = 6 \times 10^{-9} \ e^2/h$ – three orders of magnitude lower that at 4.0 T !

We find that the shapes of the curves reflect the interplay between compressibility and resistivity features. Basically, a reduced $D(ring)$ by itself can reduce the charging only directly on the ring. This accounts for the in-phase downward spikes that appear near x = ±2 µm at 4.0 T and 4.1 T, and partially at 4.2 T, in both the measured and modeled curves. In contrast, a low conductivity restricts the flow of charge across the ring, resulting in a reduced in-phase signal *everywhere* in the interior. Hence instead of a ring-shape, this phenomenon produces a disk-shaped feature.

## IV. COMPARISON TO SIMPLE MODELS

The advance modeling implies that the measured behavior arises from the reduced $\sigma_{xx}$ and $D$ of a low-compressibility ring. As a check, we now compare this result to the simple models introduced in Section II.

As the structure observed at 4.0 T is consistent with arbitrarily high $\sigma_{xx}(ring)$, it is reasonable to apply the single parallel-plate capacitor model introduced in Sec. II (A) to estimate the compressibility reduction responsible for the in-phase ring (seen as downward spikes). Taking the ring as a 10% reduction of $Q_{in}$, we apply Eq. 1 using $d$=90 nm, $h$=10 nm, and κ=12.5. We then find that the simple model yields a reduced compressibility of $D(ring) = 5 \times 10^{11}$ cm$^{-2}$eV$^{-1}$ (a 60-fold reduction compared to the zero magnetic field compressibility of $D \approx 3 \times 10^{13}$ cm$^{-2}$eV$^{-1}$). This roughly agrees with the value found by the advanced model of $2 \times 10^{11}$ cm$^{-2}$eV$^{-1}$. The agreement is surprising considering that by neglecting the effects of stray capacitance, the single capacitor model requires the tip-sample mutual capacitance to account for most of the electric field lines emanating from the sample. In contrast, the numerical routine assumes a small $C_{mut}/C_{stray}$ to calculate the effective potential without regard to the tip position. Hence, we may conclude that in this case the extracted value of ring compressibility is fairly insensitive to the details of the model.

With regard to the structure seen at 4.1-4.3 T, we can apply the simple *RC* circuit shown in Fig. 2 (c, d). Because we believe the resistance and capacitance are mostly separated (i.e., not distributed) in the real system, the *RC* model should approximately predict the correct magnitude of the signal interior to the ring; $R$ is set by the low-conductivity boundary, whereas the self-capacitance of the interior disk determines $C$. Looking at the curves qualitatively, we can conclude that at 4.2 T the system is near the out-of-phase peak. This follows from the observation that the measured $Q_{out}$ data have the maximum interior amplitude and are fairly insensitive to frequency; at this same field, the $Q_{in}$ data have the maximum sensitivity to frequency. This behavior is consistent with 30 kHz ~ $1/2\pi RC$. Using the $C$=1.0 fF for the capacitance of a 2 µm radius disk near the surface of the dielectric, we find $R\sim 6$ GΩ. This implies $\sigma_{xx}(ring) \sim 1\times 10^{-7}$ $e^2/h$, using a "number of squares" factor consistent with a 400 nm ring thickness. This compares well with $1.5\times 10^{-7}$ $e^2/h$, the best fit $\sigma_{xx}(ring)$ of the advanced model at 4.2 T.

**VI. CONCLUSION**

In summary, using Subsurface Charge Accumulation imaging, we have observed incompressible regions that evolve smoothly with magnetic field, consistent with quantum Hall effect theories considering long-range potential fluctuations in the presence of disorder. In addition to observing the spatial patterns of the signal, we have shown that by carefully comparing the charge accumulation signal to modeling, approximate values of the local compressibility and conductivity can be extracted from the measurements.

Surprisingly, we find that relatively small increments of the magnetic field lead to changes in the electrical resistivity across the incompressible strip of more than three orders of magnitude. This drastic variation is approximately consistent with recent measurements of incompressible strips formed using metal gates deposited on a sample surface. [18] Interestingly, in both cases the strip resistivity was higher as the strip moved into the region of higher density gradient, where the width of the strip is expected to be smaller [10].

**APPENDIX**

The purpose of this appendix is to give more details of the numeric calculations for our advanced modeling for charging within the 2D layer. With the imaged area of the sample defined as a $N \times N$ array of grid points, the starting point for the calculations is the charge conservation expression for each point $k$:

$$\frac{dq_k}{dt} - \sum_l I_{k,l} = 0, \quad (A1)$$

where the charge term (first term) gives the rate of change of the charge and the current term (second term) gives the total current flowing into the grid point from the surrounding points. As the current is a function of the potential of the points, it can be rewritten, $\sum_l I_{k,l} = \sum_l S_{k,l}\varphi_l$, or in matrix notation $S\vec{\varphi}$. Here $S$ is an $N^2 \times N^2$ matrix containing the conductivities $\sigma_{xx}^k$, $\sigma_{xy}^k$ between adjacent pairs of points, and $\vec{\varphi}$ is a vector with $N^2$ elements – each corresponding to the potential of a particular point $k$. In linear response, the charge term can also be expressed as a function of $\vec{\varphi}$ by constructing a capacitance matrix $C$, $d\vec{q}/dt = (i2\pi f)\vec{q} = (i2\pi f)C\vec{\varphi}$.

To find $C$, we first construct a potential matrix $P$ that considers the coulomb interaction among all pairs of points, invoking image charges to account for the bound charges at the dielectric surface. The capacitance matrix is then given by $C = P^{-1}$. In contrast to S, which is non-zero only near the diagonal, C is an $N^2 \times N^2$ matrix containing no zeros. The difference is that the conductivity of the grid can be described as a network of resistors connecting only neighboring points, whereas every point has a capacitance to every other point (and a self-capacitance). Substituting the new expressions for the two terms in Eq. (A1), we find

$$(i2\pi f)C\vec{\varphi} - S\vec{\varphi} = \dot{\vec{Q}}_{source}, \quad (A2)$$

where $\dot{\vec{Q}}_{source}$ gives the locations of the injected excitation, and is only nonzero for points along the perimeter.

Eq. (A2) describes correctly the internal potential of the grid. However we have not yet included the effects of compressibility variations. Essentially as $D$ decreases, less charge accumulates in the 2DES, which induces less charge on the tip. In other words, the mutual capacitance is reduced. But because our scheme takes $C_{mut}$ as a constant, we must account for this effect as a reduction in potential.

We include the compressibility contribution $D_k$ by altering the second term of Eq. (A2), which controls the current flowing into each point. For example, to account for a low compressibility of point $k$, we adjust the $S$ matrix so that more current flows into $k$ as a function of $\varphi_k$. Therefore to conserve current, $\varphi_k$ must be reduced. The appropriately adjusted conductivity matrix $S'$ is found by defining a second potential matrix $P'$. $P'$ is identical to $P$ except that it includes the compressibility contribution $D_k$ for the inverse self-capacitance of each grid point:

$$P'_{kk} = P_{kk} + \frac{1}{e^2 D_k}.$$

$S' = SP'C$ replaces $S$ in Eq. (A2).

Solving for $\vec{\varphi}$, we obtain

$$\vec{\varphi} = [C + (i/2\pi f)S')]^{-1} \vec{\bar{Q}}_{source}, \quad (A3)$$

where $\vec{\bar{Q}}_{source} = \vec{\dot{Q}}_{source} / i2\pi f$. Hence for a given $C$, $S'$, and $\vec{\bar{Q}}_{source}$, we can calculate the effective potential $\vec{\varphi}$ of each point of the array. Note that this is not an iterative calculation yielding an approximate solution. Rather, we arrive directly at the exact $\vec{\varphi}$ using Eq. (A3).

As the scanned area is part of a much larger sample, any calculated structure arising from proximity of the array's edges has no physical meaning. We remove such artifacts by using periodic boundary conditions. Measured features arising from significant structures beyond the scanned area can lead to discrepancies with the calculations which are considerably harder to correct. However, we emphasize that these non-local effects can only lead to broad features, and cannot produce structure at length scales ~100 nm. As the final step in the calculation, we find the charge induced on the tip, which allows direct comparison to the measurements. This is accomplished using Eq. (3) of Sec. III.

## ACKNOWLEDGEMENTS

We gratefully acknowledge L.S. Levitov for numerous key discussions and for developing essential elements of the numerical routine. We thank M. R. Melloch for providing the sample used in the experimental measurements, and A. Shytov for programming assistance. This work was supported by the Office of Naval Research, the Packard Foundation, JSEP, and the National Science Foundation DMR.